\documentclass[fleqn,usenatbib]{mnras}
\usepackage{graphicx,xspace}
\usepackage{dcolumn, xcolor}
\usepackage{bm}
\usepackage[fleqn]{amsmath}
\usepackage{amsfonts,amssymb} 
\usepackage[e]{esvect}
\usepackage{hyperref}
\usepackage{bigints}
\usepackage{xparse}
\usepackage{rotating}
\usepackage[utf8]{inputenc}
\usepackage[T1]{fontenc}
\usepackage{mathptmx}
\usepackage{ wasysym }
\usepackage{twoopt}
\usepackage[T1]{fontenc}
\usepackage{ae,aecompl,fancyvrb}

\newcommand{\tr}[1]{\ensuremath{\mathrm{tr}(\tensor{#1})}}

\NewDocumentCommand\hyp{oo}{
  \IfNoValueTF{#1}{\ensuremath{\ \mathrm{_{[]}F_{[]}}}}
     {\IfNoValueTF{#2}{\mbox{\ensuremath{\ _{[#1]}\mathrm{F}_{[]}}}}
       {\mbox{\ensuremath{\ _{[#1]}\mathrm{F}_{[#2]}}}}
}}

\NewDocumentCommand\Uka{oo}{
  \IfNoValueTF{#1}{\ensuremath{\ {\matcal{U}^{[#2]}}}}
     {\IfNoValueTF{#2}{\mbox{\ensuremath{\ ^{[#1]}\mathcal{U}}}}
       {\mbox{\ensuremath{\ ^{[#1]}\mathcal{U}^{[#2]}}}}
     }
   }
\NewDocumentCommand\uka{oo}{
  \IfNoValueTF{#1}{\ensuremath{\ {\mathrm{^{[]}u}}}}
     {\IfNoValueTF{#2}{\mbox{\ensuremath{\ ^{[#1]}\mathrm{u}}}}
       {\mbox{\ensuremath{\ ^{[#1]}\mathrm{u}}}}
}}
\NewDocumentCommand\Wka{oo}{
  \IfNoValueTF{#1}{\mbox{\ensuremath{\ \ ^{{[]}}\mathcal{W}^{[]}}}}
     {\IfNoValueTF{#2}{\mbox{\ensuremath{\ ^{[#1]}\mathcal{W}^{[]}}}}
       {\mbox{\ensuremath{\ ^{[#1]}\mathcal{W}^{[#2]}}}}
}}
\NewDocumentCommand\wka{oo}{
  \IfNoValueTF{#1}{\ensuremath{\ \mathrm{^{[]}w}}}
     {\IfNoValueTF{#2}{\mbox{\ensuremath{\ ^{[#1]}\mathrm{w}}}}
       {\mbox{\ensuremath{\ ^{[#1]}\mathrm{w}}}}
}}
\NewDocumentCommand\Vka{ooo}{
  \IfNoValueTF{#1}{\mbox{\ensuremath{\ \ ^{{[]}}\mathcal{V}^{[]}^{#3}}}}
     {\IfNoValueTF{#2}{\mbox{\ensuremath{\ ^{[#1]}\mathcal{V}^{[]}^{#3}}}}
       {\mbox{\ensuremath{\ ^{[#1]}\mathcal{V}^{[#2]}^{#3}}}}
}}\NewDocumentCommand\sinf{oo}{
  \IfNoValueTF{#1}{\ensuremath{\sin\varphi}}
     {\IfNoValueTF{#2}{\ensuremath{\sin^{#1}\varphi}}
       {\ensuremath{\sin^{#1}#2}}
}}
\NewDocumentCommand\cosf{oo}{
  \IfNoValueTF{#1}{\ensuremath{\cos\varphi}}
     {\IfNoValueTF{#2}{\ensuremath{\cos^{#1}\varphi}}
       {\ensuremath{\cos^{#1}#2}}
}}
\NewDocumentCommand\sint{oo}{
  \IfNoValueTF{#1}{\ensuremath{\sin\vartheta}}
     {\IfNoValueTF{#2}{\ensuremath{\sin^{#1}\vartheta}}
       {\ensuremath{\sin^{#1}#2}}
}}
\NewDocumentCommand\cost{oo}{
  \IfNoValueTF{#1}{\ensuremath{\cos\vartheta}}
     {\IfNoValueTF{#2}{\ensuremath{\cos^{#1}\vartheta}}
       {\ensuremath{\cos^{#1}#2}}
}}

\newcommand{\pa}{\ensuremath{_{\parallel}}}
\newcommand{\se}{\ensuremath{_{\perp}}}

\providecommand\tensor[1]{\ensuremath{\overleftrightarrow{#1}}}

\newcommand{\GA}[1]{\ensuremath{\Gamma\left(#1\right)}}

\newcommand{\md}{\ensuremath{\mathrm{d}}}
\providecommand\half{\ensuremath{\frac{1}{2}}}

\newcolumntype{R}{>{$}r<{$\ \ =}}
\newcolumntype{C}{>{$}c<{$\hspace{0.cm}}}
\newcolumntype{K}{>{$}r<{$\hspace{0.cm}}}
\newcolumntype{L}{>{\hspace{-0.1cm}$}l<{$}}
\newcolumntype{X}{>{\ $}l<{$}}

\newcommand{\led}{\textcolor{black}} 

\let\oldsqrt\sqrt
\def\sqrt{\mathpalette\DHLhksqrt}
\def\DHLhksqrt#1#2{%
\setbox0=\hbox{$#1\oldsqrt{#2\,}$}\dimen0=\ht0
\advance\dimen0-0.2\ht0
\setbox2=\hbox{\vrule height\ht0 depth -\dimen0}%
{\box0\lower0.4pt\box2}}

\allowdisplaybreaks

\title{Generalized \led{anisotropic} $\kappa$-cookbook: 2D fitting of Ulysses electron data}

\author[K. Scherer et al.]{K. Scherer,$^{1,2}$\thanks{kls@was.tp4.rub.de}
  and E. Husidic$^{1,3}$, M. Lazar,$^{1,3}$, H. Fichtner,$^{1,2}$
\\
$^{1}$Institut f\"ur Theoretische Physik, Lehrstuhl IV:
  Plasma-Astroteilchenphysik, Ruhr-Universit\"at Bochum, D-44780 Bochum,
  Germany\\
$^{2}$Research Department, Plasmas with Complex Interactions,
  Ruhr-Universit\"at Bochum, 44780 Bochum, Germany \\
$^{3}$Centre for Mathematical Plasma Astrophysics, 
      3001 Leuven Belgium
}
\pubyear{}
\date{}
\VerbatimFootnotes
\begin{document}
\label{firstpage}
\pagerange{\pageref{firstpage}--\pageref{lastpage}}
\maketitle

\begin{abstract}Observations in space plasmas reveal particle velocity distributions out of thermal equilibrium, with anisotropies (e.g., parallel drifts or/and different temperatures, $T_\parallel$ - parallel and $T_\perp$ - perpendicular, with respect to the background magnetic field), and multiple quasithermal and suprathermal populations with different properties. The recently introduced (isotropic) $\kappa$-cookbook is generalized in the present paper to cover all these cases of anisotropic and multi-component distributions reported by the observations. 
We derive general analytical expressions for the velocity moments and show that the common (bi-)Maxwellian and (bi-)$\kappa-$distributions are obtained as limiting cases of the generalized anisotropic $\kappa$-cookbook (or recipes).
Based on this generalization, a new 2D fitting procedure is introduced, with an improved level of confidence compared to the 1D fitting methods widely used to quantify the main properties of the observed distributions.
The nonlinear least-squares fit is \led{applied to electron data sets} measured by the Ulysses spacecraft confirming the existence of three different populations, a quasithermal core and two suprathermal (halo and strahl) components. In general, the best overall fit is given by the sum of a Maxwellian and two generalized $\kappa$-distributions.
\end{abstract}

\begin{keywords}plasmas,  Sun: heliosphere, solar wind, methods: data analysis 
  \end{keywords}

\section{Introduction}\label{sec:introduction}

Space plasmas such as the solar wind or planetary magnetospheres are dilute and (nearly) collisionless \citep{Marsch-2006, Heikkila-2011}, and particle velocity distributions measured in-situ reveal non-equilibrium states, both for electrons \citep{Stverak-etal-2008, Wilson-etal-2019b, Wilson-etal-2020} and ions \citep{Gloeckler-2003, Kasper-etal-2006, Lazar-etal-2012a}. The distributions exhibit typical non-Maxwellian features such as suprathermal tails due to an excess of high-energy particles \citep{Maksimovic-etal-1997, Stverak-etal-2008}, or deviations from isotropy, e.g., temperature anisotropies with respect to the magnetic field direction \citep{Marsch-2006,Stverak-etal-2008}, or field-aligned beams (e.g., electron strahls) along the magnetic field lines with speeds higher than the bulk plasma flow \led{\citep{Pierrard-etal-2001, Wilson-etal-2019a}.}

 Up to a few keV the electron velocity distributions in space plasmas reveal three distinct components \citep{Pierrard-etal-2001, Wilson-etal-2019a}: at low energies a quasi-thermal and highly dense core ($\approx$ 80\% - 90\% of the total particle number density) approximated by a Maxwellian; and two suprathermal populations, a hot, tenuous halo ($\approx$ 5\% - 10\% of the total density) enhancing suprathermal tails, which are well described by power-law functions such as the $\kappa$-distribution function (see below), and likewise hot and tenuous beams ($\approx$ 5\% - 10\% of the total density), which can also be fitted by $\kappa$-distributions, see e.g., \cite{Stverak-etal-2008,Wilson-etal-2019b, Wilson-etal-2020}. Electrons in the solar wind are described by different combinations of multi-component models, \led{i.e., dual }models to fit only the core and halo when the strahl is not prominent enough, triple models to include also the asymmetric strahl, or even a quadruple model to include two counterbeaming strahls (double strahl) observed in closed magnetic field topologies, associated to coronal loops or interplaneteraty shocks \citep{Pilipp-etal-1987c, Lazar-etal-2012, Lazar-etal-2014, Wilson-etal-2019a, Macneil-etal-2020}.

Kappa \led{(or $\kappa$-)} power laws have become a widely employed tool to describe the observed distributions in non-equilibrium distributions, either as a global fit to incorporate both the core and halo \citep{Olbert-1968,Vasyliunas-1968}, or as \led{a partial fit} to reproduce only the halo or strahl components, see the review by \cite{Pierrard-Lazar-2010} and references therein. In the present work we propose a generalized fitting model  capable to reproduce any of the aforementioned anisotropies or combinations of beams (or drifting populations) with intrinsic anisotropic temperatures, as revealed by the realistic distributions in space plasmas \led{\citep{Maksimovic-etal-2005, Stverak-etal-2008, Pilipp-etal-1987b, Pilipp-etal-1987c, Pilipp-etal-1987a}}. In kinetic studies, e.g., when exploring the stability properties of these distributions, it is important to capture their non-Maxwellian features, and thus quantify the free energy sources triggering instabilities and wave fluctuations \citep[e.g.\ ][] {Astfalk-Jenko-2016}. In the absence of collisions space plasmas are governed by the wave-particle interactions.

The anisotropic model mostly invoked is the  bi-$\kappa$-distribution function (BK) \citep{Summers-Thorne-1992, Lazar-Poedts-2009}
 \begin{align}\label{eq:sbk} 
	f_\mathrm{BK}(\vec{r},\vec{v},t) = 
	&\frac{n(\vec{r},t)}{\sqrt{\pi^{3}\,\kappa}\,\theta_\parallel\,\theta_\perp^{2}} 
	\frac{\Gamma(\kappa)}{\Gamma\left(\kappa- \frac{1}{2}\right)} \nonumber \\
	&\times \left(1 + \frac{(v_\parallel - u_{\parallel 0})^2}{\kappa\,
	\theta_\parallel^2} + \frac{v_\perp^2}{\kappa\,
	\theta_\perp^2} \right)^{-(\kappa + 1)},
\end{align}
where $n(\vec{r},t)$ is the number density (depending on location $\vec{r}$ and time $t$), $\Gamma(\mu)$ is the (complete) Gamma function, $v$ the particle speed, which is normalized to a nominal thermal speed $\theta$, $u_{\parallel 0}$ is the drift (or bulk) speed in parallel direction, and $\kappa$ is a free parameter determining the width/slope of the suprathermal tails. The symbols $\parallel$ and $\perp$ denote directions parallel and perpendicular to the background magnetic field, respectively. The BK is defined only for values of $\kappa > 3/2$, and as $\kappa \to \infty$, it approaches the standard bi-Maxwellian distribution function (BM):
\begin{align}\label{eq:bm}
	f_\mathrm{BM}(\vec{r},\vec{v},t) = 
	\frac{n(\vec{r},t)}{\sqrt{\pi^{3}}\,\theta_\parallel\,\theta_\perp^{2}}  
	 \exp\left(-\frac{(v_\parallel - u_{\parallel 0})^2}{\theta_\parallel^2} - \frac{v_\perp^2}{\theta_\perp^2} \right)\,.
\end{align}
If this BM approximates the core of the BK, it is described by the same thermal speeds $\theta_{\parallel, \perp}$,  see \cite{Lazar-etal-2015, Lazar-etal-2016} for an extended discussion motivating a definition of the kinetic temperature for $\kappa$-distributed plasmas, i.e., $\theta_{\parallel,\perp}$ independent of $\kappa$.

Despite their successful applications, standard $\kappa$-distribution models are undermined by certain unphysical implications, such as diverging higher-order velocity moments (which restricts values of the $\kappa$-parameter to always exceed some minimum limits, although the observations may suggest even lower values) and unrealistic contributions from superluminal particles \citep{Scherer-etal-2019a,Fahr-Heyl-2020}, and thus prevent a macroscopic modeling of $\kappa$-distributed plasmas. These issues have been fixed by the regularized bi-$\kappa$-distribution function (RBK) \citep{Scherer-etal-2017,Scherer-etal-2019a, Lazar-etal-2020} 
\begin{align}\label{eq:rbk}
	f_\mathrm{RBK}(\vec{r},&\vec{v},t) = 
	\frac{n(\vec{r},t)}{\sqrt{\pi^{3}\,\kappa}\,\theta_\parallel\,\theta_\perp^{2}} 
	\frac{\Gamma(\kappa)}{\Gamma\left(\kappa- \frac{1}{2}\right)} \mathcal{W}_{[0,0]} \times \nonumber \\
	&\left(1 + \frac{(v_\parallel - u_{\parallel 0})^2}{\kappa\,
	\theta_\parallel^2} + \frac{v_\perp^2}{\kappa\,
	\theta_\perp^2} \right)^{-(\kappa + 1)} 
	e^{-\xi_\parallel \frac{(v_\parallel - u_{\parallel 0})^2}{\theta_\parallel^2} - \xi_\perp \frac{v_\perp^2}{\theta_\perp^2}}\,,
\end{align}
where $\xi_{\parallel,\perp}$ are the cut-off parameters and $\mathcal{W}_{[0,0]}$ is given by
\begin{equation}
    \mathcal{W}_{[0,0]} = \left(\int\limits_{0}^{1} \md t \,
    U \left(\frac{3}{2},\frac{3 - 2\,\kappa}{2},
    \kappa[\xi_\perp^2 + (\xi_\parallel - \xi_\perp)t^2] \right) \right)^{-1}
\end{equation}
with $U(a,c,x)$ being the Kummer-U or Tricomi function (e.g., \cite{Oldham-etal-2010}). For $\xi_\parallel = \xi_\perp = 0$ the RBK turns into the BK, and if in addition $\kappa \to \infty$, \led{the BM} is recovered.

A universally accepted version would be useful to identify and differentiate physical conditions when each of these versions of $\kappa$-distributions applies (see \cite{Scherer-etal-2020} and references therein). \cite{Scherer-etal-2020}  proposed an isotropic generalized $\kappa$-distribution, named the \led{(isotropic)} $\kappa$-cookbook, and \led{exploring} the implications of various parameter setups. Here, in \led{Section~2} we introduce an extended version able to cover cases of anisotropic and multi-component distributions reported by the observations, by introducing the \led{generalized anisotropic $\kappa$-distribution function (GAK), calling it the anisotropic $\kappa$-cookbook}. All the commonly used anisotropic model distributions, i.e., the BM, the BK and the RBK, can be thus derived as particular cases of \led{the GAK}. 
We provide the reader with general analytical expressions for the velocity moments (see Appendix \ref{app:A}), when one is not only concerned about the quality of the fits to the observations, but also about the implications for macroscopic quantities when using a certain distribution function.
Most fitting procedures invoke 1D model functions such as the BM or the BK, which are used to fit 1D cuts along parallel, perpendicular or any other oblique direction of the observed distributions or individual components \citep{Pilipp-etal-1987b, Pilipp-etal-1987c, Pilipp-etal-1987a, Stverak-etal-2008, Wilson-etal-2019a}. While this is a common procedure, the accuracy in reproducing the overall shape of the distribution and unveil details (e.g., counterbeams mimicking  an excess of parallel temperature) is questionable. In \led{Section~3} we present a new 2D fitting method using the GAK with a Levenberg-Marquardt algorithm \citep{Levenberg-1944, Marquardt-1963} to minimize the nonlinear least-squares, which here is applied to selected electron data sets, measured by the SWOOPS instrument of the Ulysses spacecraft \citep{Bame-etal-1992}. The method allows us to depict electron populations and sum-up their model distributions to find the best overall fit and parameters controlling the quality of the fits. In \led{Section~4} we present the conclusions, pointing to the importance and the potential of a generalized model combined with a 2D fitting to provide realistic descriptions of the particle velocity  distributions observed in space plasmas. 

\section{The anisotropic $\kappa$-cookbook}

Similarly to the isotropic generalized $\kappa$-distribution function (GKD) in \cite{Scherer-etal-2020}, here we introduce the generalized anisotropic $\kappa$-distribution function (GAK) as
\begin{align}\label{eq:Gak}
  f_{GAK}&(\eta\pa,\eta\se,\zeta,\xi\pa,\xi\se)
  = n_{0}\,N_{{GAK}}\\\nonumber
      &\times \left(1+ \frac{(v_\parallel - u_{\parallel 0})^2}{\eta\pa\Theta\pa^{2}} +
      \frac{v\se^{2}}{\eta\se\Theta\se^{2}}\right)^{-\zeta}
      \mathrm{e}^{-\xi\pa\frac{(v_\parallel - u_{\parallel 0})^2}{\Theta\pa^{2}}-\xi\se\frac{v\se^{2}}{\Theta\se^{2}}}\,,
\end{align}
which we will call accordingly the anisotropic $\kappa$-cookbook. The 5-tuples $(\eta\pa,\eta\se,\zeta,\xi\pa,\xi\se)$, 
which define the cookbook, we will call 'recipes'. 
The integrals for the velocity moments are quite similar to those used in 
\cite{Scherer-etal-2019b, Scherer-etal-2020} and are given  in detail in Appendix~\ref{app:A}. The normalization constant $N_\mathrm{GAK}$ in Eq.~\eqref{eq:Gak} is such that $\int \mathrm{d}^3 v\, f_\mathrm{GAK}$ gives the number density $n_0$, and is obtained by 
\begin{align}
  N_\mathrm{GAK}^{-1} = &\sqrt{\pi^{3}}
  \Theta\pa\Theta\se^{2}\eta\pa^{\frac{1}{2}}\eta\se \\\nonumber
  &\times \int\limits_{0}^{1}U\left(\frac{3}{2},\frac{5}{2}-\zeta,
     (a_{2} +  (a_{1}-a_{2})t^{2}\right) \md t \,.
\end{align}

With the definition of the GAK by a tuple $(\eta_{1},\eta_{2},\zeta,\xi_{1},\xi_{2})$, 
    we can state the recipes for the bi-Maxwellian as $(1,1,0,1,1)$, for the standard bi-$\kappa$-distribution as $(\kappa,\kappa,\kappa+1,0,0)$ and for the regularized bi-$\kappa$-distribution as $(\kappa,\kappa,\kappa+1,\xi_{1},\xi_{2})$, see for further discussion of these
    distributions \cite{Scherer-etal-2019b}.
Here we will focus on the fitting of the high
resolution electron data from the SWOOPS instrument of the Ulysses
mission\footnote{\url{http://ufa.esac.esa.int/ufa/\#data}}. We used
also  hourly average electron data \citep[][same webpage]{Bame-etal-1992} and the magnetic field data \citep[][same webpage]{Balogh-etal-1992} for comparison and to determine the Alfv\'en speed.

\section{Data analysis}
\subsection{The model distributions}

\led{In order to differentiate between the anisotropic models used in the fitting procedure, we use the following notations $f_\alpha$, with $\alpha = \{a,b,c,d,e\}$ presented in Table~\ref{tab:four_distributions}, where RBK$^{*}$ is a modified version of the RBK, allowing for an anisotropic $\kappa$-parameterization, i.e., $\kappa_\perp \ne \kappa_\parallel$.}
\begin{table}\label{tab:four_distributions}
\begin{tabular}{l|l|l}
    Index & Recipe & Distribution \\ \hline
    a     & (1,1,0,1,1) & Bi-Maxwellian\\
    b     & $(\eta_{\pa},\eta_{\se},\zeta,\xi_{\pa},\xi_{\se})$ & GAK\\
    c     & $(\kappa,\kappa,\kappa+1,\xi_{\pa},\xi_{\se})$ & RBK\\
    d     & $(\kappa_{\pa},\kappa{\se},\kappa_{\pa}+\kappa_{\se},\xi_{\pa},\xi_{\se})$ & RBK$^{*}$\\
    e     & $(\kappa,\kappa,\kappa+1,0,0)$ & BK\\
\end{tabular}
\caption{Anisotropic distribution functions used in the fitting procedure in Sec.~\ref{sec:data_fitting}}
\end{table}

In general the electron data up to a few keV reveal three populations, a thermal core and \led{two suprathermal populations called halo and strahl, and model distributions can be combined as}
\begin{align}\label{eq:sum_distributions}
    f_{\alpha_{1}\alpha_{2}\alpha_{3}} \equiv
 f_{\alpha_{1}} + f_{\alpha_{2}} + f_{\alpha_{3}} \,,
 \end{align}
where $\alpha_{i} \in \alpha$ with $i= \{1,2,3\}$ \led{and} the ordering is according to the decreasing number density from the core (most dense) to the halo and strahl, though the strahl may sometimes be more dense than the halo, e.g., in the outer corona \led{\citep{Halekas-etal-2020}} or the CME (single or double) strahls \led{\citep{Skoug-etal-2000, Anderson-etal-2012}}.
\begin{align}
    n_{\alpha_{1}} > n_{\alpha_{2}} \geqslant n_{\alpha_{3}}
\end{align}
For example, the fit for the observed distribution function $f_{abc}$ can be the sum of the core fit with the recipe (1,1,0,1,1), 
and two suprathermal components, one fitted by the recipe
$(\eta_{\pa},\eta_{\se},\zeta,\xi_{\pa},\xi_{\se})$, 
and the other one by $(\kappa,\kappa,\kappa+1,\xi_{\pa},\xi_{\se})$
(with the lowest number density $n_{\alpha_{3}}$).
\led{The bi-$\kappa$-distribution function is not used explicitly in the fitting procedure, but is included in the application of the GAK and the RBK, e.g., if in the RBK $\xi_\parallel = \xi_\perp = 0$.}

The velocity moments for a sum of distribution functions are already discussed in \cite{Scherer-etal-2020}:
\begin{align}
n_{\alpha_{1}\alpha_{2}\alpha_{3}} &= \sum\limits_{i=1}^{3}n_{\alpha_i}
    \\\label{eq:ucms}
  \vec{u}_{cms} &\equiv \vec{u}_{\alpha_{1}\alpha_{2}\alpha_{3}} =   \frac{\sum\limits_{i=1}^{3}n_{i}\vec{u}_{\alpha_i}}
    {\sum\limits_{i=1}^{3}n_{\alpha_i}} =(u_{\pa,cms},\vec{u}_{\se})^{T}\\
      \tensor{P} &= \sum\limits_{i=1}^{3}\int f_{i}(\vec{v})
      \left[\vec{v}\otimes\vec{v}
     + (\vec{u}-\vec{u}_{i})\otimes(\vec{u}-\vec{u}_{i})
      \right] \md^{3}v\\\label{eq:Ppa}
      P_{\pa} & \equiv \sum\limits_{i=1}^{3} P_{\alpha_{i},11}
      + \sum\limits_{i=1}^{3} n_{\alpha_i}(u_{\pa,cms}-u_{\pa,i})^2\\
     P_{\se} & \equiv \sum\limits_{i=1}^{3} P_{\alpha_{i},22}\\\label{eq:Pse}
      \vec{H} &= \half \sum\limits_{i=1}^{3} \left(2\tensor{P}_{\alpha_{i}} + \tr{P}_{\alpha_{i}} - n_{\alpha_{i}}U_{\alpha_{i}}^{2}\right)\vec{U}_{\alpha{i}}\\\label{eq:Hpa}
 \led{H_{\pa}} &\led{= \sum\limits_{i=1}^{3}\left(3 P_{\alpha_{i},11} + 2 P_{\alpha_{i},22} - n_{\alpha_{i}} U_{\pa,\alpha_{i}}^2\right)U_{\pa,\alpha_{i}}}\\
      H_{\se} &= 0 
\end{align}
with $\vec{U}_{\alpha{i}} = \vec{u}_{cms} - \vec{u}_{\alpha{i}}$ , where $\vec{u}_{cms}$ is the center of mass velocity. $\tensor{P}$ denotes the thermal pressure tensor and $\vec{H}$ is the heat flux vector.
It is always assumed that the perpendicular drift speed is zero (see
discussion below) and thus $\vec{u}_{\se,\alpha{i}}\equiv 0$
and $\vec{u}_{\se,cms}=0$,
and we are only left with the parallel components in the center of
mass velocity as well as in the heat flux $\vec{H}$.
The above moments are all normalized to the electron mass.

 \begin{figure*}
     \centering
     \includegraphics[width=0.45\textwidth]{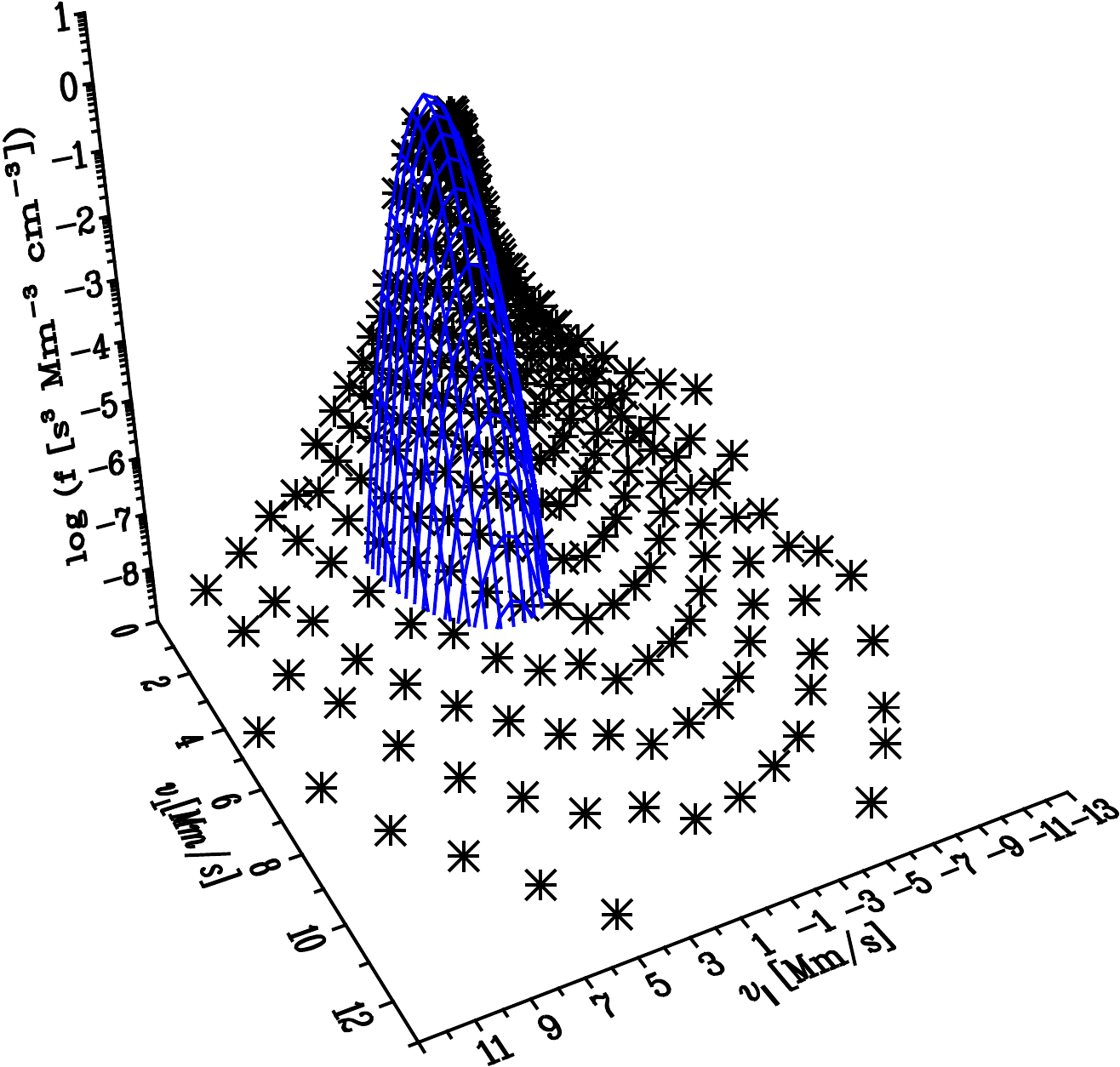}    
     \includegraphics[width=0.45\textwidth]{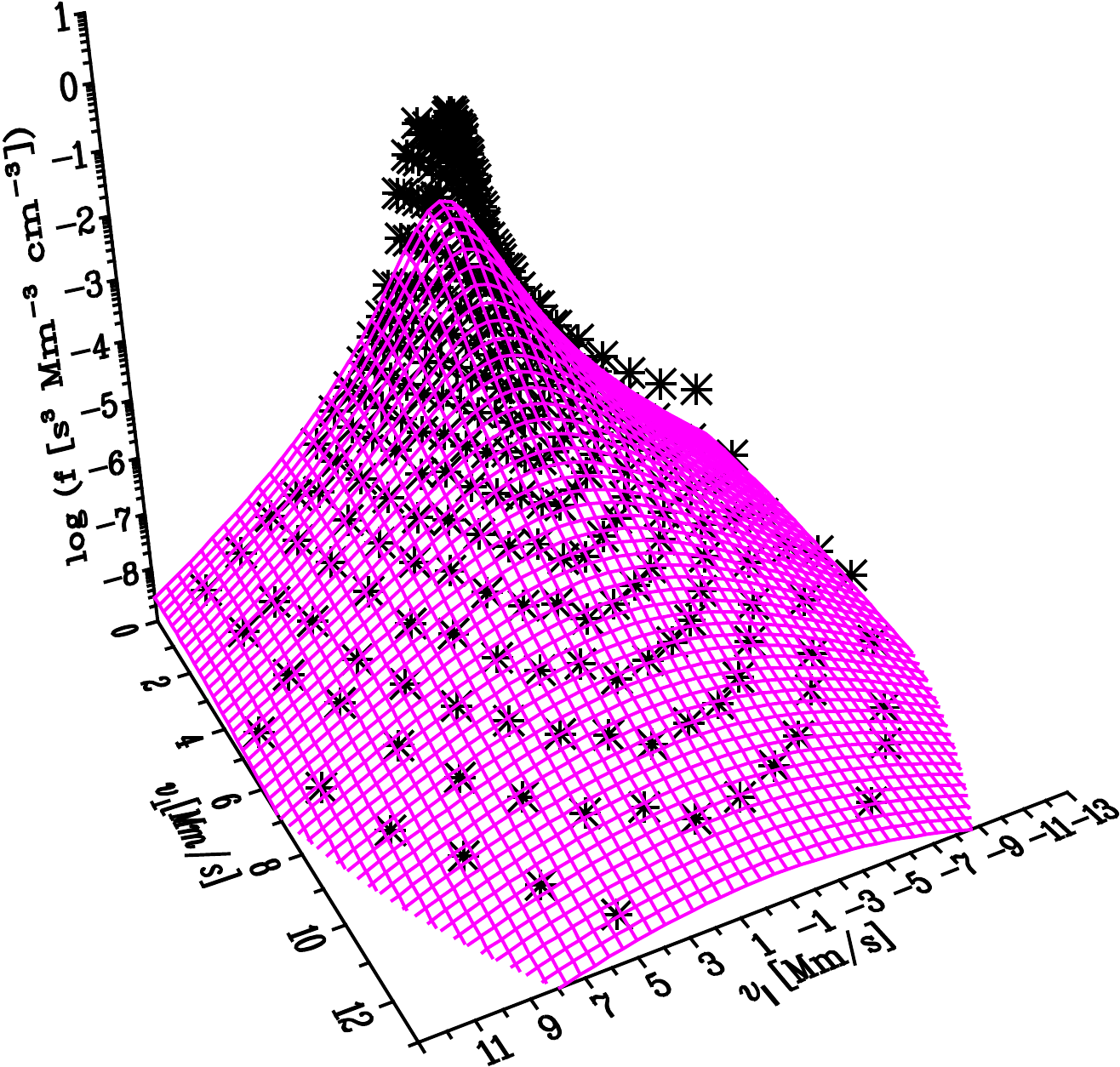}\\
     \includegraphics[width=0.45\textwidth]{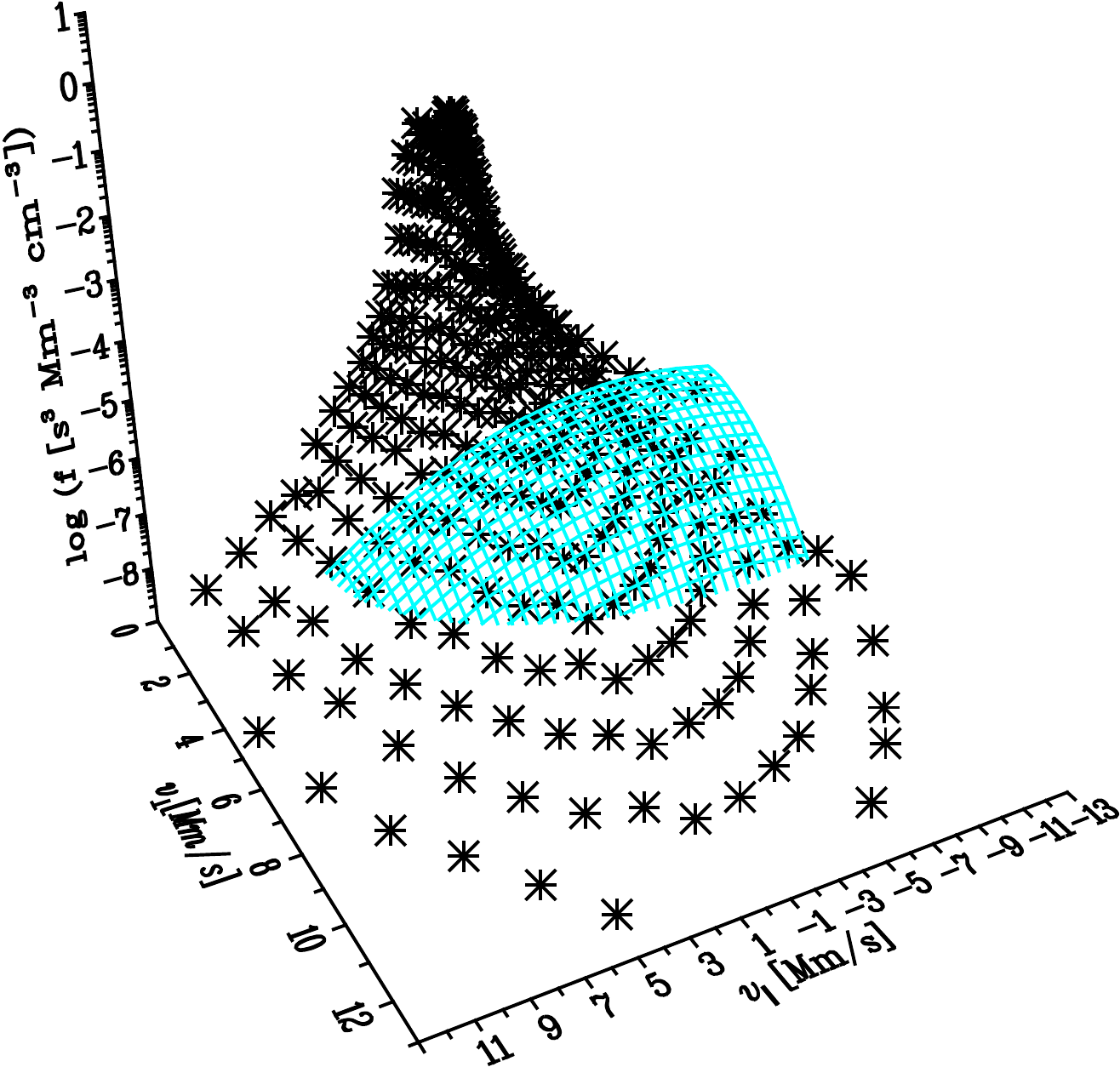}    
     \includegraphics[width=0.45\textwidth]{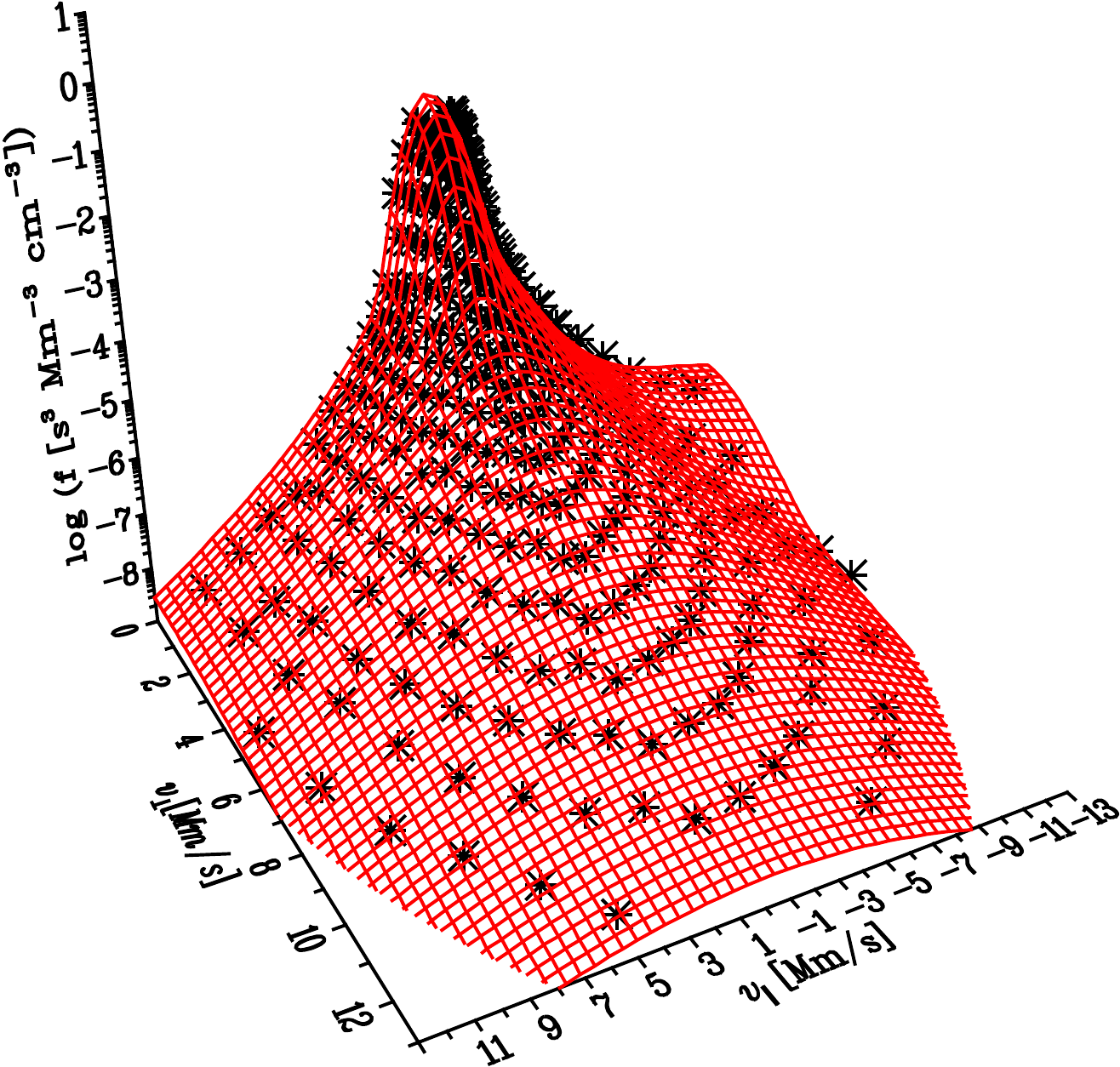}
     \caption{3D representation of the data (2002, doy 14,22:09:58) by a Maxwellian, a GAK and  a Maxwellian: On the x-axis the
       parallel speed $v_{\pa}$
       and on the y-axis the perpendicular speed $v_{\se}$
       in units of [Mm/s] are given, while on the logarithmic z-axis
       the distribution function in [{cm$^{-3}$(Mm/s)$^{-3}$}] is
       shown. The black crosses represent the data, while the colored grids
       show the following fits: top left panel a Maxwellian fit of the core, 
       top right a GAK fit of the suprathermal part, bottom 
       left a shifted Maxwellian fit of the strahl, bottom right the sum of all 
       three model distributions.
       One can see that the data are best fitted with the sum of
       three distribution functions. The fits are done simultaneously, and the single distributions are shown in the plots above.}
     \label{fig:3d}
 \end{figure*}

In Fig.~\ref{fig:3d} we show the data (black crosses) and the simultaneously fitted
distribution functions (colored grids). In the upper left panel only the core is
fitted with a Maxwellian  (Eq.~\ref{eq:bm}), in the upper right panel only the first
suprathermal part is fitted with a GAK (Eq.~\ref{eq:Gak}), in the lower left panel
only the second suprathermal part is fitted with a Maxwelian (Eq.~\ref{eq:bm}), and finally in the lower right
panel the fit in form of the sum of all three model distribution is presented. 

 From Fig.~\ref{fig:3d} it becomes evident that the data consists of
 three populations: a core and two suprathermal populations, where the hotter population is called strahl 
 and the less hot population constitutes the halo. The best fit is then obtained with the sum of three model distributions.  We fitted all three distribution functions at once, and the single distributions shown in Fig.\ref{fig:3d} are then the single distributions taken from the fit.

 \subsection{Data fitting}\label{sec:data_fitting}

 We use a nonlinear least-squares fit based on the Marquardt-Levenberg 
 \citep{Levenberg-1944, Marquardt-1963}
 algorithm \led{to find the best (combination of) model distribution(s)}. During that procedure
 we fitted all possible combinations of only one (global) fit
 $f_{\alpha_{i}}$, dual fits $f_{\alpha_{i}\alpha_{j}}$, or triple fits combining three 
models $f_{\alpha_{i}\alpha_{j}\alpha_{k}}$ as shown in Eq.~\eqref{eq:sum_distributions}
 (for more details see Appendix~\ref{app:C}).
 f we fit more than one distribution the fit to all distributions is always done simultaneously. This prevents us from deciding which part of the data belongs to a given distribution. For testing we fitted first a Maxwellian (for $v<3$ Mm/s, then using that initial values two distributions, and with the same procedure at least three distributions. There were only negligible differences to the full fit, and thus we always do a full fit. 
 
 For the accuracy of fitting we defined the relative error
 \begin{align}
     E_{i} = \frac{|f_{fit,i}-f_{obs,i}|}{f_{obs,i}}
 \end{align}
 for each data point $i$,  and the mean relative error $<E>$, its standard deviation $\sigma$ and the $R^{2}$ value as
 \begin{align}\label{eq:mean}
     <E> &= \frac{1}{N} \sum\limits_{i=1}^{N} E_i \,, \\\label{eq:sigma}
     \sigma &=\sqrt{\frac{1}{N-1} \sum\limits_{i=1}^{N} (<E> -
              E_i)^2} \,, \\\label{eq:rsqr}
   R^{2} & = 1 - \frac{ (\sum\limits_{i=1}^{N} E_i-<E>)^{2}}{\sum\limits_{i=1}^{N} E_i^{2}} \,.
 \end{align}
The values for the mean relative error $<E>$ and the standard deviation $\sigma$ are required to be as small as possible for a satisfying fit. Here, by direct visual inspection (comparison of the different fits) we define values below 0.3 for $<E>$ and $\sigma$ as a ``good'' fit, and thus only such fits are considered in the following. However, in order to evaluate the goodness of the fits not only visually, but to have a quantitative measure, we use the $R^{2}$ values. Once $<E>$ and $\sigma$ are below 0.3, the highest $R^{2}$ value determines the best fit, while the limit $R^{2}=1$ is the idealized best fit (i.e., $<E> \to 0$ and $\sigma\to0$). 

In the present analysis we compare two data sets arbitrarily chosen from the year 2002: day 288, 00:08:17 (hh:mm:ss), and day 365, 00:13:45.  Table~\ref{tab:msr} presents all tested model distributions for these data sets, the values obtained for $<E>$ and $\sigma$ with the corresponding $R^{2}$ values. 
It is to be observed that for day 365 (right column) not $f_\mathrm{adb}$ gives the best fit (despite the lowest $<E>$ and $\sigma$ values), but $f_\mathrm{aba}$ due to the highest $R^2$ value.

\begin{table}
 \scalebox{0.85}{ \begin{tabular}{l|lll||l|lll||}
    \multicolumn{4}{c}{day 228} & \multicolumn{4}{c}{day 365}\\
  $f_{\alpha_{i}\alpha_{j}\alpha_{k}}$ &$<E>$&$\sigma$&$R^{2}$&    $f_{\alpha_{i}\alpha_{j}\alpha_{k}}$ &$<E>$&$\sigma$&$R^{2}$\\
aaa & 0.28 & 0.24 & 0.57 &    aaa &  0.21 &      0.18 &  0.58\\
aba & 0.23 & 0.20 & 0.56 & \bf   aba &\bf  0.28 & \bf     0.19 &\bf  0.69\\
\bf abb &\bf 0.27 & \bf 0.18 & \bf 0.70 & -&-&-&-\\
   -&-&-&-&                   abc &  0.16 &  0.15 &  0.55\\
aca & 0.29 & 0.27 & 0.54 &    aca &  0.19 &  0.19 &  0.51\\
   -&-&-&-&                   acb &  0.17 &  0.16 &  0.54\\
add & 0.29 & 0.24 & 0.61 &       -&-&-&-  \\
   -&-&-&-&                   ada &  0.17 &      0.17 &  0.50\\
   -&-&-&-&                   adb &  0.15 &      0.14 &  0.54\\  
  \end{tabular}}
\caption{The mean average error $<E>$, the standard deviation $\sigma$ 
        and the $R^2$.
        The model distribution which gives the best fit is marked in
        bold \led{for the particular day}. The $-$ sign \led{indicates} that these values of $<E>$
        and $\sigma$ are above 0.3. All other combinations above 0.3 are left out.}
  \label{tab:msr}
\end{table}

\begin{figure*}
     \centering
     \includegraphics[width=0.45\textwidth]{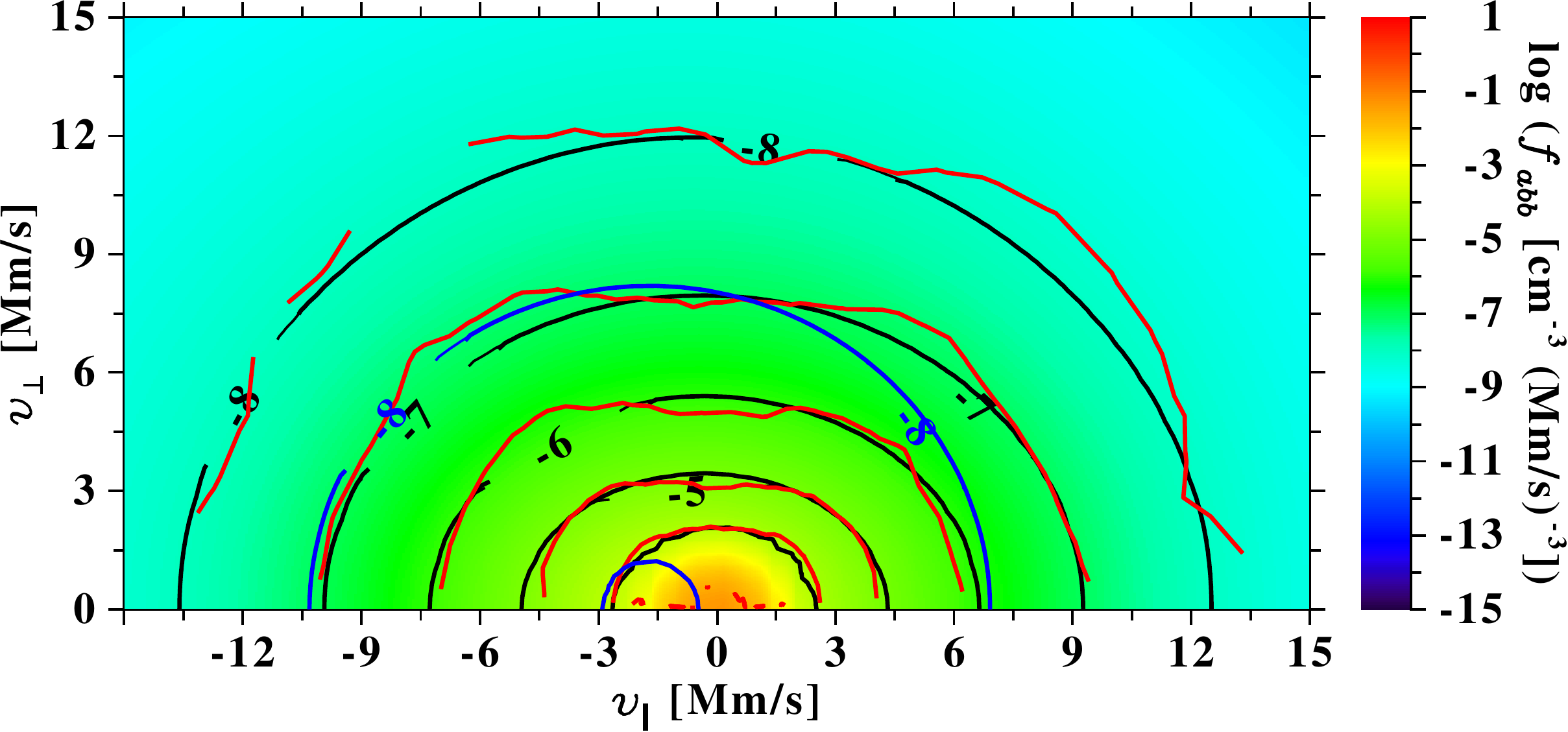}    
     \includegraphics[width=0.45\textwidth]{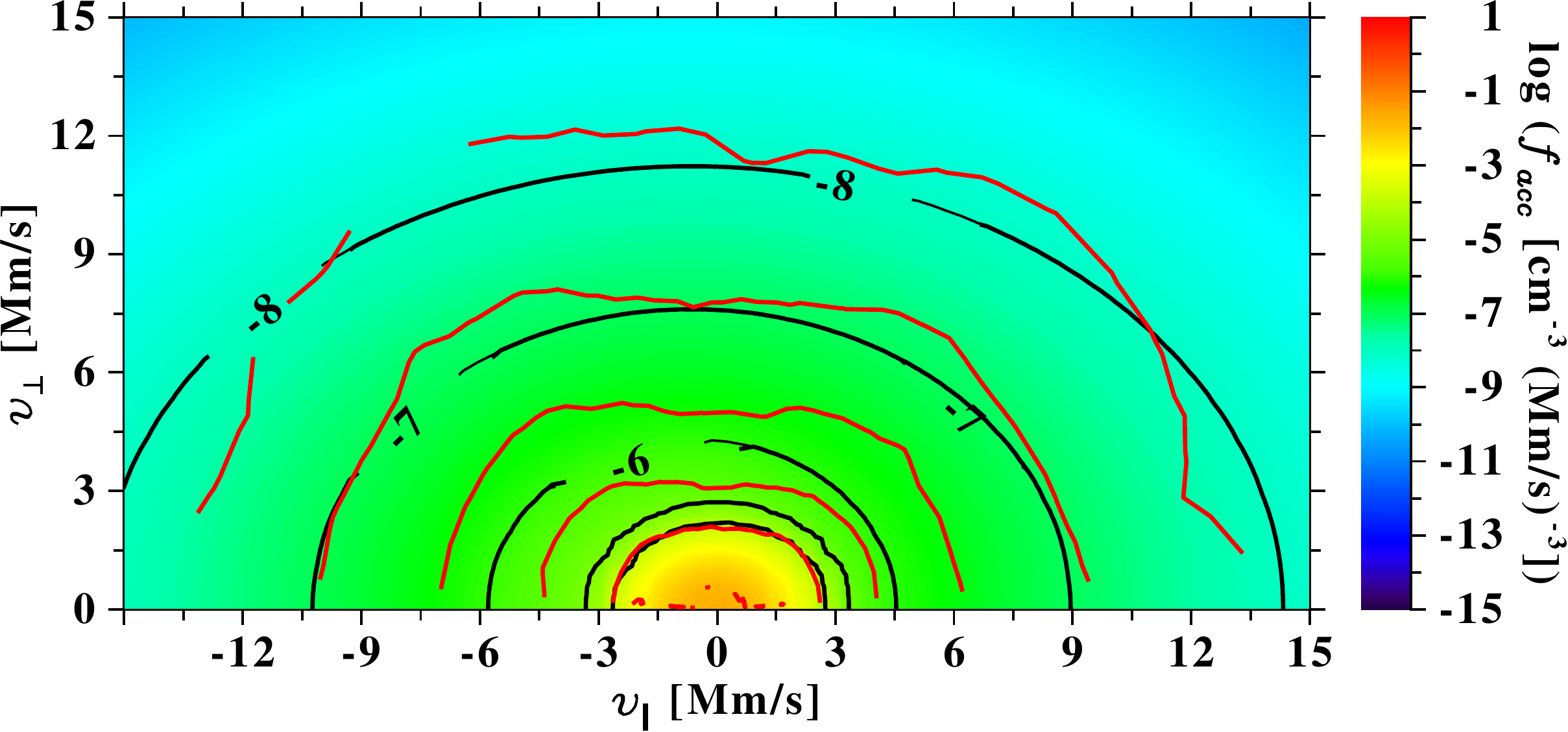}\\
     \includegraphics[width=0.45\textwidth]{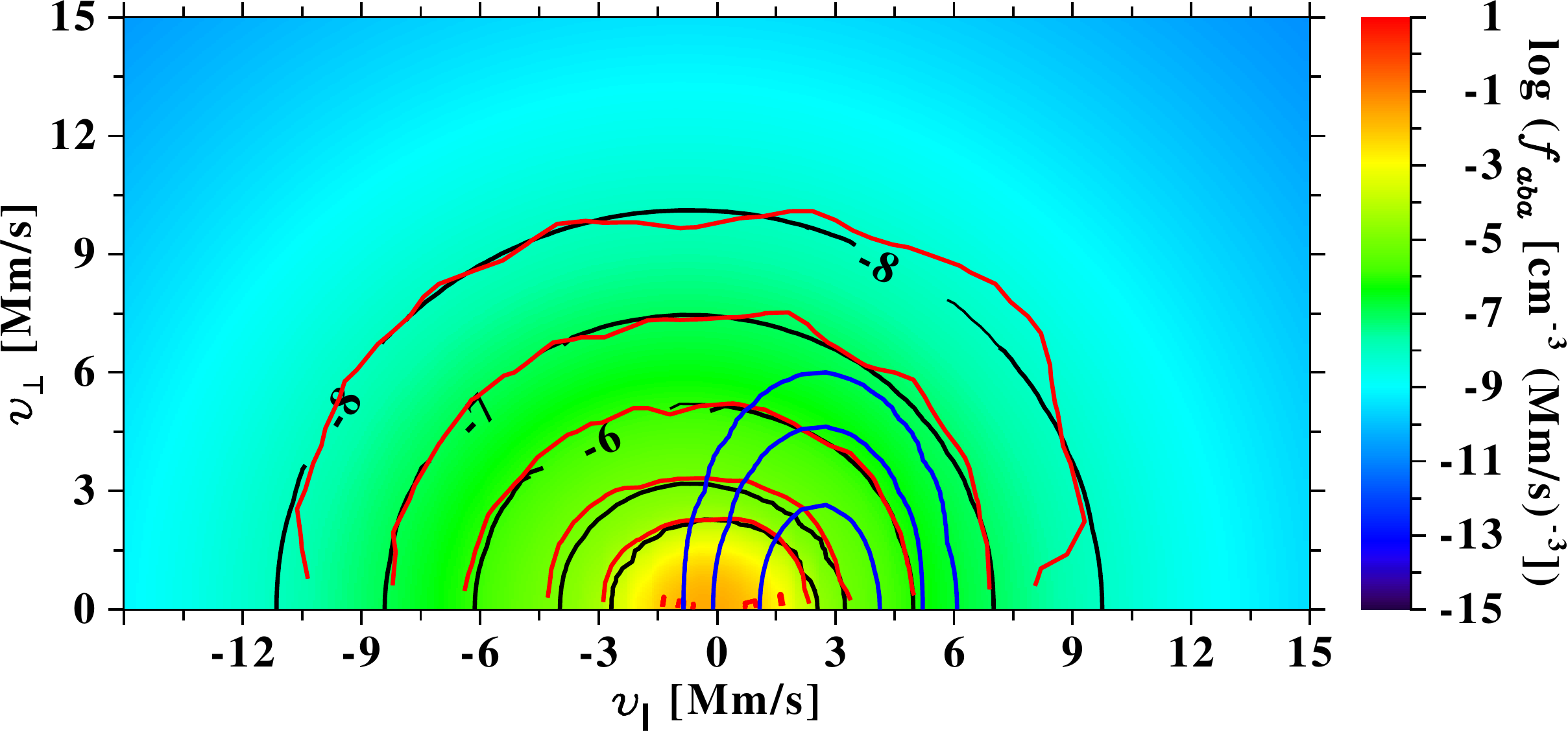}    
     \includegraphics[width=0.45\textwidth]{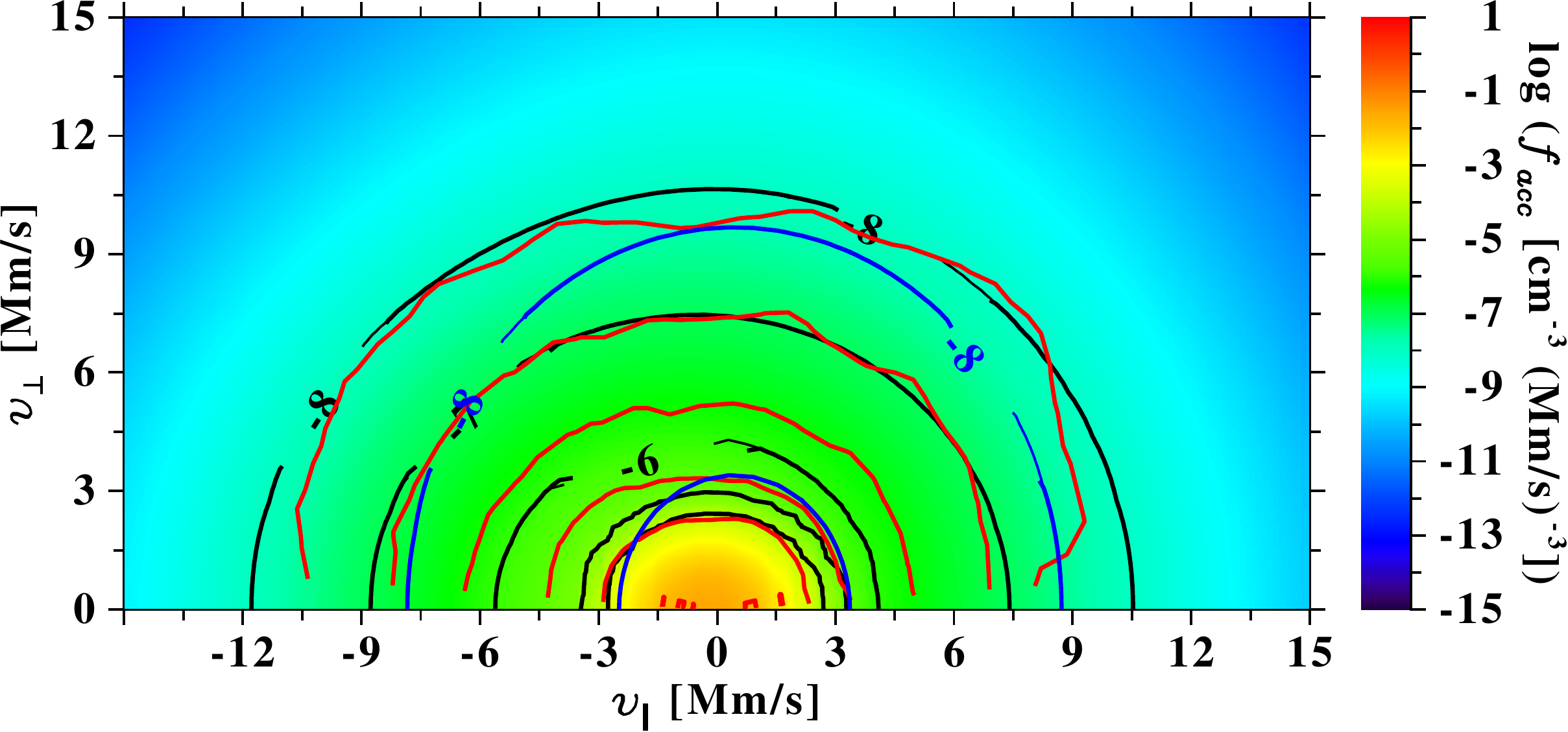}
     \caption{Contour plots of the fitted distribution functions. The
       upper panels show two fitted models at the date 2002, day 288, 00:08:17:
       the upper left panel shows the best model $f_{abb}$
       with $<E>=0.27$,
       $\sigma=0.18$,
       and $R^{2}=0.70$,
       while in the upper right panel a ``bad'' model $f_{acc}$
       is shown with $<E>=0.5$, $\sigma=0.37$, and $R^{2}=0.65$. In
       the lower panels two fitted models at the date 2002, day 365,
       00:13:45 are shown:
     the lower left panel shows the best model $f_{aba}$
       with $<E>=0.28$,
       $\sigma=0.19$,
       and $R^{2}=0.69$,
       while on the lower right panel a ``bad'' model $f_{acc}$
       is shown with $<E>=0.81$, $\sigma=0.14$, and $R^{2}=0.97$. The
       bad fits are excluded because of the large $<E>$ or $\sigma$
       values \led{(above our defined threshold of 0.3, see text for details)}. The black contour lines are those of the model fit,
       while the red contours are those from the data.  The blue contour lines show the levels of the third distributions, which are in the above case too small to contribute substantially to those of the total distribution function. }
     \label{fig:1}
 \end{figure*}

In order to show differences of accuracy, in
Fig.~\ref{fig:1} we have compared the best fits from Table~\ref{tab:msr} with another bad fit, arbitrarily chosen. In this figure both electron data sets are used. The black contour lines show the model fits, while the red contour lines show that of the data. For a good fit the black and red contour lines should coincide. For a good fit (the top and bottom left panels, respectively), these contours are close to each other, while in the bad fits (the top and bottom right panels, respectively), they markedly differ. Thus our choice
first excludes all the $<E>$ and $\sigma$ values which are larger than
0.3, and than \led{chooses} from the remaining models that for which the $R^{2}$ value
is the largest as a good choice. Nevertheless, one should keep in mind
that the nonlinear least-squares fit may find a local minimum, which gives
not the best fit. Therefore, we have tested different initial values, and in rare cases, this leads to better fitting models. The presented fits are robust against different initial values, and the best model is in all tests the same with the same fitted values (up to rounding errors). 

In Table~\ref{tab:fit} we show the best fit values for the two discussed
cases. With help of the formulas \eqref{eq:ucms}, \eqref{eq:Ppa}, \eqref{eq:Pse},
\eqref{eq:Hpa}, and \eqref{eq:mom} we can calculate the center of mass
speed, the pressure and the heat flow using the fitted values. The
results are given in Table~\ref{tab:mom} together with the  integrated SWOOPS electron data \citep[see][]{Bame-etal-1992}, which give the isotropic moment for the core and halo as well as the magnetic field and the
resulting Alfv\'en speed. To make the fitted anisotropic temperatures
compatible to the  isotropic temperatures, we
estimated the corresponding isotropic pressure as
$P_{iso}=P_{\pa}+2P_{\se}$, and from there the isotropic
temperatures. One should keep in mind that the ion data data are
averaged over 1 hour, while the  electron data are given
in 2 min intervals (with a much longer cadence time of 3-17 min.)

\begin{table*}
\scalebox{0.75}{
  \begin{tabular}{ll|lll|lll|ll|l}
 day &$f_{\alpha_{i}\alpha_{j}\alpha_{k}}$   & $n_{i}$ [cm$^{-3}$] & $\Theta_{\parallel,i}$
                                           [Mm/s] & $\Theta_{\perp,i}$
                                                    [Mm/s]
  & $\eta_{\parallel,i}\kappa$ & $\eta_{\perp,i}$
                          &$\zeta$ 
        & $\xi_{\parallel,i}$ & $\xi_{\perp,i}$ &
                                                  $u_{\parallel,i}$[Mm/s]\\
    \hline
  288 &$\alpha_i=a$ &    $ 1.20 \cdot 10^{-1}$&$  8.08 \cdot 10^{-1}$&$  7.50 \cdot 10^{-1}$& &&&&&$  1.05 \cdot 10^{-1}$\\
&$\alpha_j=b$ &$  1.42 \cdot 10^{-3}$&$  2.37 \cdot 10^{-1}$&$  1.94 \cdot 10^{-1}$&$  6.59 \cdot 10^{-2}$&$  4.80 \cdot 10^{-2}$&$  5.01 \cdot 10^{-1}$&$  1.36 \cdot 10^{-3}$&$  9.86 \cdot 10^{-4}$&$ -5.25 \cdot 10^{-1}$\\
&$\alpha_k=b$ &$  9.68 \cdot 10^{-3}$&$  1.17 \cdot 10^{1}$&$  1.00 \cdot 10^{1}$&$  3.98 \cdot 10^{-1}$&$  3.08 \cdot 10^{-1}$&$  1.21 \cdot 10^{1}$&$  4.51 \cdot 10^{-3}$&$  1.12 \cdot 10^{-3}$&$ -1.72 \cdot 10^{-1}$\\
\hline
365 & $\alpha_i=a$  &$  1.34 \cdot 10^{-1}$&$  1.02 \cdot 10^{0}$&$  9.23 \cdot 10^{-1}$& &&&&&$ -4.81 \cdot 10^{-2}$\\
&$\alpha_j=b$ &$  5.63 \cdot 10^{-3}$&$  2.34 \cdot 10^{0}$&$  2.30 \cdot 10^{0}$&$  3.31 \cdot 10^{0}$&$  3.11 \cdot 10^{0}$&$  4.58 \cdot 10^{0}$&$  9.19 \cdot 10^{-4}$&$  5.88 \cdot 10^{-5}$&$ -6.97 \cdot 10^{-1}$\\
&$\alpha_k=a$ &$  1.47 \cdot 10^{-4}$&$  1.45 \cdot 10^{0}$&$  2.51 \cdot 10^{0}$& &&&&&$  2.61 \cdot 10^{0}$\\
  \end{tabular}}
  \caption{Best fit parameters.}
  \label{tab:fit}
\end{table*}

\begin{table*}
\scalebox{0.75}{  \begin{tabular}{ll|lll|l|lll|l|ll}
day & function& $n_{tot}$\,[cm$^{-3}$]&$n_{core}$\,[cm$^{-3}$] &$n_{halo}$\,[cm$^{-3}$]&$u_{cms}$\,[Mm/s]&$T_{tot}$\,[K]&$T_{core}$\,[K]&$T_{halo}$\,[K]&$H_{\pa}$\,[gs$^{-3}$]&$|\vec{B}|$\,[nT]&$v_{A}$\,[Mm/s]\\
 &&&&&$u_{SW}$\,[Mm/s]&$T_{ion}$\,[K]&&&&&\\
\hline
288 & abb &$ 1.7\cdot10^{-1}$ & $2.7\cdot10^{-2}$ & $1.2\cdot10^{-6}$
          &$7.8\cdot10^{-2}$
&$1.7\cdot10^{5}$&$1.6\cdot10^{5}$&$3.1\cdot10^{5}$
          &$-7.6\cdot10^{-5}$\\
&isotropic & $1.5\cdot10^{-1}$& $1.2\cdot10^{-1}$& $2.3\cdot10^{-2}$
&&
$1.2\cdot10^{5}$& $4.4\cdot10^{4}$& $5.4\cdot10^{6}$ &&
$7.7\cdot10^{-1}$ &$4.2\cdot10^{-2}$\\
&ion & $1.4\cdot10^{-1}$&&&$5.1\cdot10^{-1}$&$5.8\cdot10^{5}$&&&&&\\
                    \hline
365& aba& $1.4\cdot10^{-1}$ & $1.3\cdot10^{-1}$& $5.6\cdot10^{-3}$
& $-7.1\cdot10^{-1}$
& $7.3\cdot10^{5}$& $3.1\cdot10^{5}$& $4.6\cdot10^{5}$
& $1.2\cdot10^{-5}$&&\\
&isotropic & $3.8\cdot10^{-1}$& $3.7\cdot10^{-1}$& $1.2\cdot10^{-2}$
&& $6.3\cdot10^{5}$& $4.7\cdot10^{5}$& $5.2\cdot10^{6}$&
&$9.38\cdot10^{-1}$& $5.1\cdot10^{-2}$\\
&ion& $4.1\cdot10^{-1}$&&&$4.2\cdot10^{-2}$&$6.5\cdot10^{5}$&&&&&\\
                    \hline
                  \end{tabular}}
  \caption{The moments for the fitted anisotropic model distributions
    and those from the isotropic data. The magnetic field
    and the Alvf\'en speed are also displayed. Additionally, the rows
    indicated by ``ion'' give the proton number densities, speeds and
    temperatures from the SWOOPS instrument. The temperature is calculated from the pressure $P_{tot}=P_{\pa}+2P_{\se}$ using the ideal gas law for the core, halo and the total pressure (using Eq.~\ref{eq:Ppa}). The position of the
    Ulysses spacecraft on day 288 is at $r=4.2$\,au, $\vartheta=29.88^{o}$, 
    $\varphi=18.09^{o}$, and on day 365 at $r=4.5$\,au, $\vartheta=24.00^{o}$, 
    $\varphi=8.94^{o}$.\label{tab:mom}} 
\end{table*}

From Table~\ref{tab:fit} one can also see that fits of the anisotropic distributions give similar results as those for the isotropic data and the ion number density. The match is not perfect, possibly for various reasons: 1. The data recording does not occur at the same time, or
2. the data are averaged over different time intervals, or 3. there may be major differences between isotropic and anisotropic distribution functions. However, the number densities and temperatures are in the same range. The difference also affects the third fitting model,
which can be much hotter than expected (at day 288 $T_{\pa,3}=1.6\cdot10^{8}$\,[K] and $T_{\se,3}= 1.2\cdot10^{8}$\,[K] and at day 365
$T_{\pa,3}= 5.4\cdot10^{6}$\,[K] and $T_{\se,3}= 2.1\cdot10^{6}$\,[K] ). 

\begin{figure}
\centering
 \includegraphics[width=0.45\textwidth]{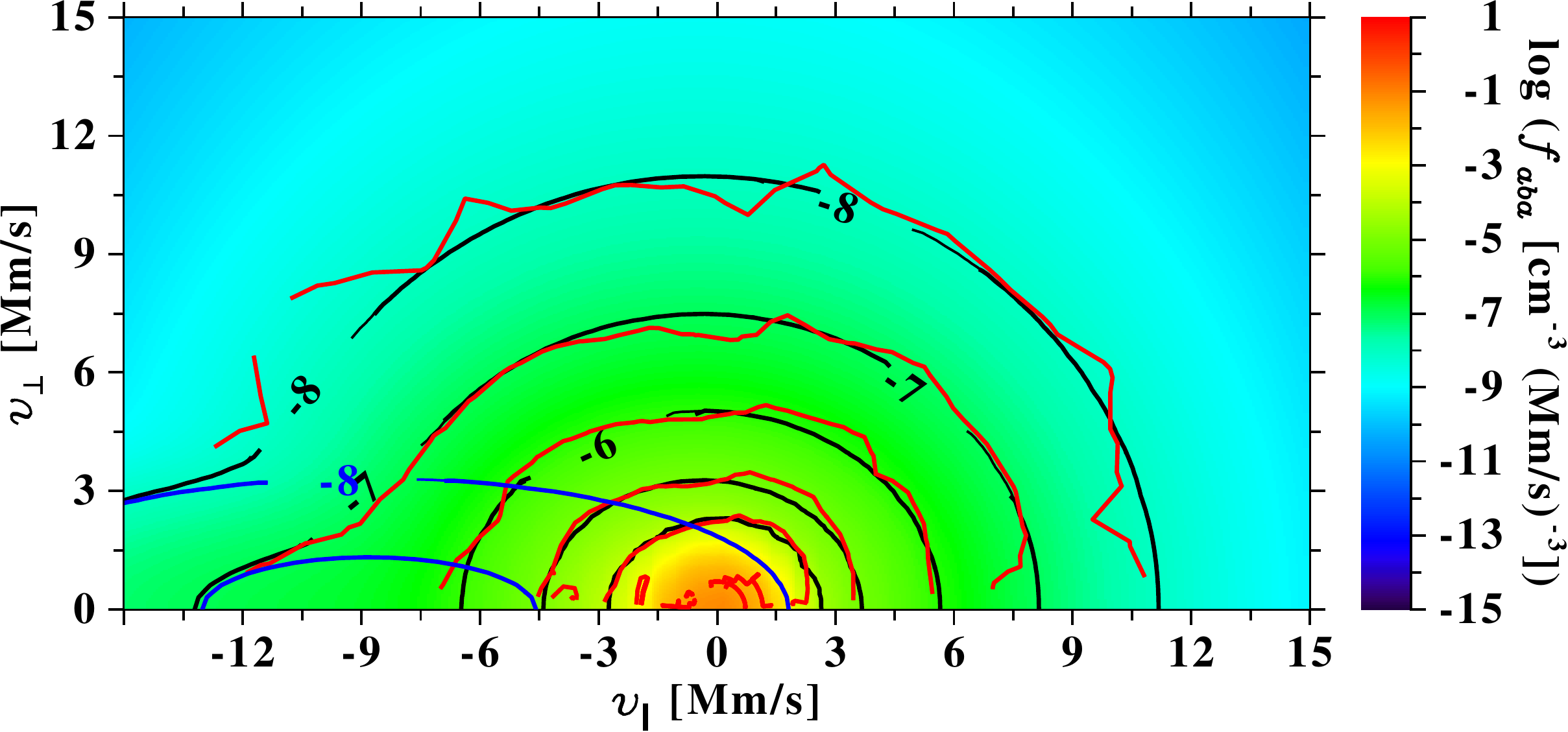}   
 \caption{Finally we show here the contour lines for  distributions shown in Fig~\ref{fig:3d}. The best fit is given by the a Bi-Maxwellian, a GAK, and again a shifted Bi-Maxwellian. The data are taken at 2002, doy 14, 22:09:52 see also \citep{Lazar-etal-2014} for the same data set. It can be seen that the red contours for the observed data scatter a lot and our quality criteria fail. We thus have chosen the fit with the lowest mean error (0.33). The black contours show the fit with all three distributions, and the blue contours only the fit for the strahl.}\label{fig:3}
 \end{figure}

In Fig.~\ref{fig:3} we show an event including a distinct strahl. Unfortunately, the data scatter a lot, so that the fit has larger mean errors and standard deviations. Therefore, we have chosen the lowest values for $<E>$ and $\sigma$ (neglecting $R^2$) for the best fit. The number density, temperature and drift speed  for the three fitted distributions (Maxwellian, GAK, Maxwellian)  are shown in Table~\ref{tab:strahl}. For convenience we have also added the temperature anisotropy in Table~\ref{tab:strahl}. It can be seen that the core is a "cold" isotropic distribution, the halo is a hot moderate anisotropic distribution, and the strahl is an even hotter highly anisotropic distribution. A discussion of the physical implications, like heat transfer, will go beyond the scope of the recent discussion and will be object to further work.

\begin{table}
    \begin{tabular}{l|rrrrr}
        $f_{a}$ &$n$[cm$^-3$]& $T_{\pa}$[K] & $T_{\se}$[K] & $u_{\pa}$[Mm/s] & $A=T_{\se}/T_{\pa}$\\
        \hline
         a &$2.7\cdot10^{-1}$&   23194 &  22792&  0.14 & 0.99\\
         b &$1.7\cdot10^{-2}$&  270454 & 172650&  -0.36 &0.64\\
         a &$2.2\cdot10^{-5}$& 6584576 & 135541&  -8.80&0.02\\
    \end{tabular}
    \caption{The moments for the distribution function shown in Fig.~\ref{fig:3}. For the GAK the remaining parameter are $\eta_{\pa}=1.3$, $\eta_{\se}=1.2$, $\zeta = 3.1$, $\xi_{\pa}=2.6\cdot10^{-2}$, and $\xi_{\se}= 9.5\cdot10^{-5}$ }
    \label{tab:strahl}
\end{table}

To fit the distribution in Fig.~\ref{fig:3} our above described quality parameter for the goodness of the fit failed, because of the data scattering. In that case one has either to weaken the conditions or to check the fits by hand. Nevertheless, for quite smooth data our quality assessment is sufficient. 
In the data set where the data points seem to scatter, an additional provided error estimate for each data point would be helpful, because then one can use a weighted fitting procedure.

 \section{Conclusions}
In hot and dilute plasmas from space, studies of fundamental processes, e.g., particle heating, instabilities, etc., require laborious kinetic approaches conditioned by the velocity distributions of plasma particles and their basic features, e.g, anisotropies with respect to the magnetic field combined with suprathermal high-energy tails described by the $\kappa$-distribution functions. Recently, this empirical model has been regularized in order to provide a well-defined macroscopic description of $\kappa$-distributed plasmas as fluids \citep{Scherer-etal-2017,Scherer-etal-2019a,Lazar-etal-2020}. \cite{Scherer-etal-2020} have also introduced the $\kappa$-cookbook,  a generalization such that the multitude of forms of $\kappa$-distributions but also Maxwellian limits invoked in the literature are recovered as particular cases or recipes. In the present paper we have advanced this notion by including the anisotropic distributions, often reported by the observations, in a new generalized anisotropic $\kappa$-cookbook (GAK), and provided the reader with analytical expressions of the integrals for the velocity moments.
 
In addition, this generalized model has been applied to describe the velocity distributions measured in-situ in the solar wind, by introducing a new fitting method with two-dimensional (2D) models. We demonstrated the quality of the method for samples of high resolution electron data sets provided by Ulysses missions. It turned out that the moments of our best fit distribution models are in a good agreement with the isotropic electron and ion data. The best fit may include various combinations of the GAK recipes, which can vary from a data set to another. These can be directly related to different properties of plasma species, or even their components, e.g., electron core, halo or strahl populations, conditioned by the solar activity, heliocentric distance and solar wind processes. Numerical reasons can also be invoked, because for example, for large $\kappa$-values the distribution function becomes a Maxwellian, which numerically can lead to a $\kappa-$model distribution instead of a Maxwellian model distribution. From a fitting point of view these differences are marginal, but from a theoretical point of view this can cause unwanted complications (for example, with Maxwellian distributions analytic solutions are often possible, while for non-Maxwellian they are not). These results promise extended applications in the observations and modeling of the observed distributions and their dependence with distance, latitude and longitude.

 \section{Data availability statement}
 The high
resolution electron data are from the SWOOPS instrument of the Ulysses
mission: \url{http://ufa.esac.esa.int/ufa/\#data}. We used
also  hourly average electron data \citep[][same webpage]{Bame-etal-1992} and the magnetic field data \citep[][same webpage]{Balogh-etal-1992} 
 
\bibliographystyle{aa}
\bibliography{test}

\appendix

\section{The integrals}\label{app:A}

The integrals for the velocity moments $M^{\lambda,\mu}$ of order $\lambda, \mu$ of the anistropic cookbook
(Eq.~\ref{eq:Gak}) have the form (neglecting the factors $n_{0}$ and $N_\mathrm{GAK}$ and having already taken into account the drift speed)
\begin{align}
  M^{\lambda,\mu} &=
  \int\limits_{0}^{2\pi}\int\limits_{-\infty}^{\infty}\int\limits_{0}^{\infty}
  f_{Gak}(\eta\pa,\eta\se,\zeta,\xi\pa,\xi\se)
                    v\pa^{\lambda}v\se^{\mu+1} \md v\pa \md v\se \md \varphi\\\nonumber
                  &=\int\limits_{0}^{2\pi}\int\limits_{-\infty}^{\infty}\int\limits_{0}^{\infty}
           \left(1+ \frac{v\pa^{2}}{\eta\pa\Theta\pa^{2}} +
      \frac{v\se^{2}}{\eta\se\Theta\se^{2}}\right)^{-\zeta}
                    \mathrm{e}^{-\xi\pa\frac{v\pa^{2}}{\Theta\pa^{2}}-\xi\se\frac{v\se^{2}}{\Theta\se^{2}}}\\\nonumber  
        &\hspace{3cm} \times v\pa^{\lambda}v\se^{\mu+1} \md v\pa \md v\se
          \md \varphi    \\\nonumber
  & = 2\pi  \Theta\pa^{1+\lambda}\Theta\se^{2+\mu}\eta\pa^{\frac{\lambda+1}{2}}\eta\se^{\frac{\mu+2}{2}}
    \int\limits_{-\infty}^{\infty}\int\limits_{0}^{\infty}\left(1+w\pa^{2}+w^{2}\se\right)^{-\zeta}\\\nonumber
& \hspace{2cm} \times \mathrm{e}^{-\xi\pa\eta\pa\Theta\pa^{2}w\pa^{2}-\xi\se\eta\se\Theta\se^{2}w\se^{2}}
w\pa^{\lambda}w\se^{\mu+1} \md w\pa \md w\se
  \\\nonumber
  &= 2\pi
    \Theta\pa^{1+\lambda}\Theta\se^{2+\mu}\eta\pa^{\frac{\lambda+1}{2}}\eta\se^{\frac{\mu+2}{2}}
     \int\limits_{0}^{\pi}\int\limits_{0}^{\infty}
  r^{\lambda+\mu+2}\left(1+r^{2}\right)^{-\zeta}\\\nonumber
  & \hspace{2cm} \times \mathrm{e}^{-r^{2}(a_{2} +
    (a_{1}-a_{2})\cos^{2}\vartheta)} \sint^{\mu+1}\cost^{\lambda}\md r
    \md \vartheta\,,
\end{align}
where we have first replaced
\begin{align}
  v\pa =  \sqrt{\eta\pa} \Theta\pa w\pa\,, \qquad v\se = \sqrt{\eta\se}\,,
  \Theta\se w\se
\end{align}
and next
\begin{align}
  w\pa = r\cos\vartheta\,, \qquad w\se = r \sin\vartheta \,, \qquad r = \sqrt{w\pa^{2}+w\se^{2}}\,,
\end{align}
and used the shorthand notation
\begin{align}
  a_{1} = \xi\pa\eta\pa \,, \qquad a_{2}= \xi\se\eta\se \,.
\end{align}
Next we use $z=r^{2}$ and $t=\cos\vartheta$ to obtain 
\begin{align}
   M^{\lambda,\mu} =& 2\pi
    \Theta\pa^{1+\lambda}\Theta\se^{2+\mu}\eta\pa^{\frac{\lambda+1}{2}}\eta\se^{\frac{\mu+2}{2}}
                     \int\limits_{0}^{1}\int\limits_{0}^{\infty}
                     z^{\frac{\lambda+\mu+1}{2}}\left(1+z\right)^{-\zeta}\\\nonumber
  &\times \mathrm{e}^{-z (a_{2} +
    (a_{1}-a_{2})t^{2})} (1-t^{2})^{\frac{\mu}{2}}t^{\lambda}\md z
    \md t\\\nonumber
    &= 2\pi
      \Theta\pa^{1+\lambda}\Theta\se^{2+\mu}\eta\pa^{\frac{\lambda+1}{2}}\eta\se^{\frac{\mu+2}{2}}
      \GA{\frac{3+\lambda+\mu}{2}} \\\nonumber
  &
    \times \int\limits_{0}^{1}U\left(\frac{3+\lambda+\mu}{2},\frac{5+\lambda+\mu}{2}-\zeta,
     a_{2} +  (a_{1}-a_{2})t^{2}\right)\\\nonumber
&   \times (1-t^{2})^{\frac{\mu}{2}}t^{\lambda}\md t \,.
\end{align}
For the normalization constant we thus find ($\lambda=0, \mu=0$):
\begin{align}
  N_\mathrm{GAK}^{-1} = &\sqrt{\pi^{3}}
  \Theta\pa\Theta\se^{2}\eta\pa^{\frac{1}{2}}\eta\se \\\nonumber
  &\times \int\limits_{0}^{1}U\left(\frac{3}{2},\frac{5}{2}-\zeta,
     (a_{2} +  (a_{1}-a_{2})t^{2}\right) \md t \,.
\end{align}
By introducing the shorthand notations
\begin{align}
  \Wka[\lambda,\mu]&(\eta\pa.\eta\se,\zeta,\xi\pa,\xi\se) \equiv
                     \Wka[\lambda,\mu] \equiv\\\nonumber   
                   & \frac{ \int\limits_{0}^{1}U\left(\frac{3+\lambda+\mu}{2},\frac{5+\lambda+\mu}{2}-\zeta,
     a_{2} +  (a_{1}-a_{2})t^{2}\right)
  (1-t^{2})^{\frac{\mu}{2}}t^{\lambda}\md t}
  {\int\limits_{0}^{1}U\left(\frac{3}{2},\frac{5}{2}-\zeta,a_{2} +  (a_{1}-a_{2})t^{2}\right) \md t}
\end{align}
and
\begin{align}
  \Wka[] \equiv \frac{1}
   {\int\limits_{0}^{1}U\left(\frac{3}{2},\frac{5}{2}-\zeta,a_{2} +  (a_{1}-a_{2})t^{2}\right) \md t} \,,
\end{align}
we can write the moments as
\begin{align}\label{eq:mom}
  M^{\lambda,\mu} &= \frac{2}{\sqrt{\pi}} n_{0}
                      \Theta\pa^{\lambda}\Theta\se^{\mu}\eta\pa^{\frac{\lambda}{2}}\eta\se^{\frac{\mu}{2}}
                      \GA{\frac{3+\lambda+\mu}{2}} 
                     \Wka[\lambda,\mu] \,.
\end{align}
The distribution function is then
\begin{align}
  f_\mathrm{GAK} =& \frac{n_{0}}{\sqrt{\pi^{3}}
  \Theta\pa\Theta\se^{2}\eta\pa^{\frac{1}{2}}\eta\se}\Wka[] \\\nonumber
&\left(1+ \frac{v\pa^{2}}{\eta\pa\Theta\pa^{2}} +
      \frac{v\se^{2}}{\eta\se\Theta\se^{2}}\right)^{-\zeta}
      \mathrm{e}^{-\xi\pa\frac{v\pa^{2}}{\Theta\pa^{2}}-\xi\se\frac{v\se^{2}}{\Theta\se^{2}}}\,.
\end{align}

\begin{table*}
\begin{tabular}{l|rrr|rrr|rr|rr}
  \hline
  $\vec{c}(1:9,i)$ & $n_{i}$ [cm$^{-3}$] & $\Theta_{\parallel,i}$
                                           [Mm/s] & $\Theta_{\perp,i}$
                                                    [Mm/s]
  & $\eta_{\parallel,i}$ & $\eta_{\perp,i}$
                          &$\zeta$ 
        & $\xi_{\parallel,i}$ & $\xi_{\perp,i}$ & $u_{\parallel,i}$
                                                  [Mm/s]\\ 
 1& $1\cdot10^{-1}$& 1.0& 1.0 & 1.5 & 1.5& 2.5 & $1\cdot10^{-2}$ &
                                                                 $1\cdot10^{-3}$ &$5\cdot10^{-2}$ \\
 2& $1\cdot10^{-3}$& 1.5& 1.5 & 1.5 &1.5 &2.5  & $1\cdot10^{-2}$ &
                                                                 $1\cdot10^{-3}$ &$5\cdot10^{-2}$ \\
 3& $1\cdot10^{-7}$& 2.0& 2.0 & 1.5 &1.5 &2.5  & $1\cdot10^{-2}$ &
                                                                 $1\cdot10^{-3}$ &$5\cdot10^{-2}$\\
  \hline
\label{tab:1}
\end{tabular}
\caption{The initial parameter set.} \label{tab:initial_parameters}
\end{table*}
\section{Fitting procedure}\label{app:C}
\subsection{Data and normalization}

The electron high-resolution data from the Ulysses SWOOPS instrument are
given for 20 energy channels ($E_{kin}$):
1.69, 2.35, 3.25, 4.51, 6.26, 8.65, 12.1, 16.8, 23.2, 31.9, 43.9,
60.6, 84.0, 116, 163, 226, 312, 429, 591, 815\,[eV], and for 20
angle bins $\varphi$
in 9$^\circ$ steps in the range from 0$^\circ$ to 180$^\circ$, which gives us a 20$\times$20 array.

The first angle bin is along the magnetic field line. 
For the speed $v$ we use the classical energy-velocity relation
\begin{align}
    \half m_e v^2 = E_{kin}\,,
\end{align}
where $m_\mathrm{e}$ is the electron mass, and $v$ is the speed.
The speed can be decomposed into a parallel and a perpendicular component in the following way:
\begin{align}
    v = v_{\parallel}\cos\varphi + v_{\perp} \sin\varphi \,.
\end{align}
Thus, with the above angles, $v_{\pa}$
has values in the interval [$-15, \ldots, 15$]\,Mm, 
and $v_{\se}$ in the interval [$0,\ldots, 15$]\,Mm.

The distribution function is normalized to [s$^3$/cm$^3$/Mm$^3$].
The nonlinear least-squares fit is done by a Levenbergh-Marquardt
algorithm from the minpack-package \citep{Elzhov-etal-2016}. 

We performed a couple of tests, where we defined an initial function
from the intial dataset, i.e., those data which are used as best
guess.
We fit always the logarithm of the data and the model distributions, because otherwise only the core is well fitted. The reason is that the least square fit minimizes the total distances between the data points and the model. Thus in fitting the original functions  the distances of the smaller values of the distribution do not contribute much to the total distance. For a logarithmic fit the distances on all scales contribute in a similar manner to the total distances.

\subsection{Input parameter}
The array $\vec{c}(1:9,1:3)$ which is used as initial array for fitting, e.g., the
assumed ``best'' initial values is given in Table~\ref{tab:initial_parameters}.

\bsp
\label{lastpage}

\end{document}